\documentclass[conference]{IEEEtran}
\usepackage{cite}

\ifCLASSINFOpdf
\else
\fi
\usepackage{url}

\usepackage[section,frozencache,cachedir=.]{minted}
\usepackage{minted}
\usepackage{listings}
\usepackage{adjustbox}
\usepackage{longtable}
\usepackage[framemethod=TikZ]{mdframed}
\usepackage{subcaption}
\usepackage{graphicx}
\usepackage{hyperref}
\hypersetup{                    
  colorlinks,
  linkcolor={green!80!black},
  citecolor={red!70!black},
  urlcolor={blue!70!black},
  breaklinks=true
}

\newmintedfile[cppcode]{cpp}{
style=vs,
linenos=true,
breaklines,
frame=single,
breakafter=d, 
fontsize=\small,
numbersep=2pt
}

\usepackage{xcolor}

\newcommand{\mei}[1]{{#1}}
\newcommand{\added}[1]{#1}

\begin{document}
%
\title{Measuring the Runtime Performance of \added{C++} Code Written by Humans using GitHub Copilot}
%
%
%



\author{\IEEEauthorblockN{Daniel Erhabor}
\IEEEauthorblockA{\textit{University of Waterloo} \\
Waterloo, Canada \\
derhabor@uwaterloo.ca}
\and
\IEEEauthorblockN{Sreeharsha Udayashankar}
\IEEEauthorblockA{\textit{University of Waterloo} \\
Waterloo, Canada \\
s2udayas@uwaterloo.ca}
\and
\IEEEauthorblockN{Meiyappan Nagappan}
\IEEEauthorblockA{\textit{University of Waterloo} \\
Waterloo, Canada \\
m2nagapp@uwaterloo.ca}
\and
\IEEEauthorblockN{Samer Al-Kiswany}
\IEEEauthorblockA{\textit{University of Waterloo} \\
Waterloo, Canada \\
alkiswany@uwaterloo.ca}
}

    

%
%

\maketitle

\begin{abstract}
GitHub Copilot is an artificially intelligent programming assistant used by many developers. While a few studies have evaluated the security risks of using Copilot, there has not been any study to show if it aids developers in producing code with better runtime performance. We evaluate the runtime performance of \added{C++} code produced when developers use GitHub Copilot versus when they do not. To this end, we conducted a user study with 32 participants where each participant solved two C++ programming problems, one with Copilot and the other without it and measured the runtime performance of the participants' solutions on our test data. Our results suggest that using Copilot may produce \added{C++} code with (statistically significant) slower runtime performance.
\end{abstract}

%
\IEEEpeerreviewmaketitle


\section{Introduction}

Advances in natural language processing and deep learning have resulted in large language models (LLMs) that can generate code from free-form text. With this, language models like GPT-3 \cite{gpt3} have been extended to what Xu et al. \cite{NL2Code} have termed Natural-Language-to-Code (NL2Code) generators. Notably, Open AI's extension of the GPT-3 language model, Codex \cite{codex}, and the production-ready product derived from it, GitHub Copilot \cite{copilot}, are popular examples of NL2Code tools in use today. \added{In a recent StackOverflow survey, 44\% of developers state that they use LLM-based tools in their development process already, and 26\% plan to use such tools soon~\cite{SOSurvey}. While some studies show that developers may have a positive experience using GitHub Copilot \cite{expVsExp}, others show that it could generate potentially vulnerable code \cite{asleepKeyboard}.}

\added{We present the first-ever evaluation of Copilot from a runtime performance perspective in systems programming. We focus on runtime performance as it is critically important in large-scale systems. Google notes that a few additional seconds of page load latency can increase customer bounce rates by 90\% \cite{google_study}. Amazon reports that 100 milliseconds of latency cost them millions of dollars in revenue \cite{gigaspacesAmazonFound}. Each millisecond of additional latency costs financial firms \$100 million every year \cite{bigdata_finance}. Thus, large-scale systems designed to maximize performance measure and report metrics such as their tail latencies and throughputs \cite{dynamo_amazon, bigtable_google}.}

We conducted the first user-based study on Copilot to evaluate the runtime performance of the C++ code generated when developers use it. With the results from our study, we answer the following research questions:
\begin{itemize}
  \item[] \textbf{RQ0}: Does using Copilot influence program correctness?
  \item[] \textbf{RQ1}: Is there a runtime performance difference in \added{C++} code when using GitHub Copilot?
  \item[] \textbf{RQ2}: Do Copilot's suggestions sway developers towards or away from \added{C++} code with faster runtime performance?
\end{itemize}

To answer these questions, we conducted a user study involving 32 participants with systems programming experience.  Each participant solved two programming problems in C++; one was solved with Copilot and the other was solved without it. The problems were related to I/O operations and concurrent programming. We selected problems related to these two domains as they directly impact the code runtime. We compared the runtime performance of Copilot-aided solutions against Copilot-unaided solutions, obtained survey responses from participants after they completed the study, and analyzed the video recordings of participants solving the problems. 

Our findings indicate that using Copilot resulted in \added{C++} code with (statistically significant) slower runtime performance. Specifically, Copilot-unaided solutions were 26\% faster than Copilot-aided solutions on average for the I/O-related problem and 15\% faster for the concurrent programming problem. Our expert solutions to the problems had up to $6\times$ faster runtime performance compared to the average Copilot-aided solution. Additionally, Copilot's aid tended to tilt developers towards code with slower runtime performance. Finally, as expected, higher developer experience and familiarity with the C++ programming language were correlated with faster runtime performance.

The rest of this paper is organized as follows: We provide background related to GitHub Copilot and related work in Section~\ref{section:related}. The process of creating the problems solved by the participants and the rationale behind choosing the problems is described in Section~\ref{section:problems}. Our model solutions are elaborated in Section~\ref{section:solutions}. A summary of the participant recruitment process and the participants is described in Section~\ref{section:participants}. We present the experiment design in detail in Section~\ref{section:experimentDesign} where we cover the tasks that participants solved, how tasks were split across the participants, and the rationale behind it. We analyze and discuss the results of our study, answering our research questions in Section~\ref{section:evaluation}. 
\added{We include a discussion on the participants and their familiarity with the problems in Section~\ref{section:discussion}.}
Penultimately, we discuss the threats to validity of our study in Section~\ref{section:rational}.
Finally, in Section~\ref{section:conclusion}, we discuss the takeaways and potential future directions. 

\begin{figure*}
    \centering
    \includegraphics[width=0.75\textwidth]{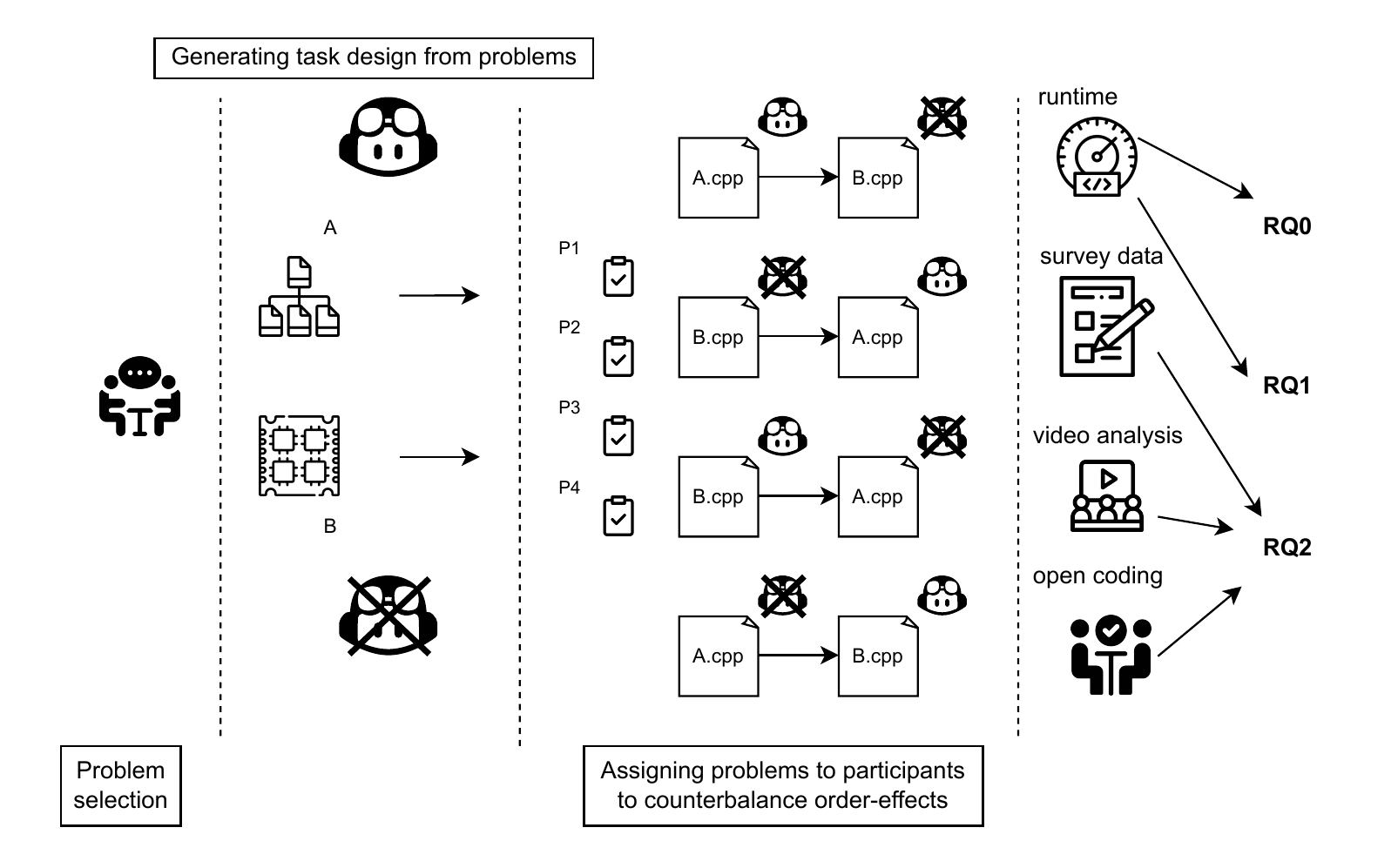}
    \caption{\added{Overview of Methodology}}
    \label{fig:methodologyDiagram}
\end{figure*}


\section{Background and Related Work}
\label{section:related}
GitHub Copilot, the production-ready tool based on the Codex model by Open AI, can be used as a Visual Studio Code extension to suggest code snippets to users. Users can receive suggestions by starting to write code or by writing comments; either way, Copilot will suggest some snippets \cite{copilot}. 


\subsection{Related Work}

\textbf{Runtime performance of code generated by Copilot or ChatGPT.} A recent study \cite{copilot_code_quality_promise_2022} analyzes the correctness and runtime performance of solutions produced by Copilot using problems from the HumanEval\cite{humaneval} dataset. They focus on comparing the multiple solutions suggested by Copilot. Doderlein et al. \cite{copilot_temperature} and Elnashar et al. \cite{chatgpt_prompt_vanderbilt} analyze the impact of prompt engineering on the runtime performance of solutions generated by Copilot and ChatGPT respectively. Mastropaolo et al. \cite{copilot_robustness} analyze the solutions generated by Copilot for different semantically equivalent task descriptions. Nascimento et al. \cite{nascimento2023comparing} evaluate the runtime performance of code produced by ChatGPT \cite{openai2023gpt4} for various LeetCode problems to their human-written solutions provided within LeetCode.  
\mei{All of the past studies focus on the solutions generated by Copilot or ChatGPT on their own and do not look at the scenario where humans are using such tools to help write code. In our study we examine code written by humans who use Copilot. This is a key difference between past work and our work.}

\textbf{Security.} Several studies within previous literature examine the security aspects of solutions generated by Copilot. One of the earliest such studies by Pearce et al. \cite{asleepKeyboard} sought to understand how often suggestions from Copilot were vulnerable to security attacks and the contexts which made Copilot suggest vulnerable code. To achieve this, they prompted Copilot to generate code in scenarios where the resulting solutions could be either vulnerable or secure. 40\% of the programs produced in these scenarios were discovered to be vulnerable. 



A study by Sandoval et al. \cite{asleepKeyboardUserStudy} assesses the security of code written by student programmers when assisted by an NL2Code assistant (OpenAI’s \lstinline{code-cushman-001} model) like Copilot. They conducted a between-subjects study with 58 computer science students where participants were tasked with implementing operations of a Singly-Linked List in C. Contrary to the previous study ~\cite{asleepKeyboard}, their results showed that Copilot had no conclusive impact on security. 


Asare et al. \cite{asare2023github} use a previously curated set of common C/C++ vulnerabilities from human developers \cite{vulnerabilities_dataset} to assess whether Copilot introduces similar vulnerabilities when presented with the same scenarios. They conclude that while Copilot is susceptible to introducing a few previously seen vulnerabilities, it fares better than human developers in a majority of the cases.


\textbf{Other factors influencing runtime performance.} Numerous studies examine the impact of other parameters, such as software refactoring \cite{software_refactoring_traini} and specific code changes \cite{peass_code_changes} on the runtime performance of open-source software repositories. These are orthogonal to our paper as we focus on evaluating the code generated by GitHub CoPilot.

\textbf{CoPilot for alternate problem domains.} Drori et al. \cite{mit_linear_solving} evaluate the effectiveness of Copilot when generating programmatic solutions to university-level linear algebra problems. Tang et al. \cite{statistics_solving} use Copilot to tackle university-level probability and statistics problems.

Dakhel et al. \cite{copilot_asset_liability} report the correctness ratio of solutions generated by Copilot for fundamental algorithmic problems. In addition, they compare Copilot's solutions against student submissions for 5 Python programming assignments. They found that while Copilot often generates "buggy" code, its repair costs are less than those of similar "buggy" student solutions. While they report the optimality of Copilot's solutions to the algorithmic problems, they do not report the runtime performance of it's solutions to the Python assignments or compare them against human submissions.

Imai et al. \cite{copilot_human_pair_prog} compare Copilot against human pair programming by having participants develop a text-based minesweeper game in Python. Nguyen et.al \cite{github_leetcode} evaluate Copilot using 33 randomly chosen questions from LeetCode, primarily focusing on solution correctness and comprehensibility. Liu et al. \cite{chatgpt_refining} characterize the correctness and maintainability of ChatGPT's solutions for 2000 programming tasks. \added{Choudhuri et al.~\cite{Choudhuri2024} look at the benefits and challenges faced by students when using ChatGPT for software engineering tasks. While this study does compare participants who used other resources with students who used ChatGPT, they did not look at the runtime performance of code.}  

Sobania et al. \cite{copilot_vs_genetic} compared Copilot's solutions against programs synthesised using genetic programming. They found that genetic programming models are more expensive to train and can sometimes result in solutions which aren't easily comprehensible for humans. They primarily examine the correctness of solutions generated using Copilot and not their runtime performance.


\mei{\textbf{Experiment with Humans using Copilot:} Unlike a majority of past studies\cite{mit_linear_solving, statistics_solving, copilot_human_pair_prog, github_leetcode, copilot_vs_genetic}  that focus on the solutions generated by Copilot on its own we focus on the scenario where humans are using such tools to help write code. Copilot was never meant to work without a human, at least to date. Therefore, these related studies do not examine Copilot in its intended environment and do not analyze the impact that such tools can have on software developers. Our study on the other hand is a more realistic experimental setting of how humans will use these tools. Finally, studies using Copilot without humans are limited to a set of simpler problems that can be auto-generated in full by Copilot. Thus, we are also able to examine more complex problems. Neither of our problems can be solved by Copilot with just a prompt and without human intervention. }

\begin{mdframed}[linewidth=2pt, linecolor=black, roundcorner=10pt, backgroundcolor = gray!20]
To the best of our knowledge, we are the first study to carry out a controlled experiment of the runtime performance of code written by humans working in tandem with Copilot. 
\end{mdframed}



\section{Programming Problems Solved by Participants}
\label{section:problems}

Following in the same vein as Pearson et al.~\cite{asleepKeyboard}, we provided \textit{incomplete code} for participants to implement as a solution to a given problem i.e., we provided code stubs and accompanying documentation for the solutions participants were asked to implement. We call the stubs \textit{problems} throughout this paper. These problems were provided to participants as a CPP file containing the function declaration, the unimplemented function definition that participants were expected to implement, i.e., the primary function, initialization functions and sanity checks to verify correctness. A main function was also provided as an entry point to call the initialization functions, the primary function, and the sanity checks in the appropriate order.

\subsection{Problem selection}
We chose two problem domains for our programming problems; file-system operations and multi-threaded programming. We chose these areas because problems in those domains directly impact application runtime performance. With file I/O operations accounting for about 30\% - 80\% of interactions in networked file systems~\cite{largeNFSWork}, there is a need for file system operations to be fast on storage devices~\cite{optimizingIO}. Choosing a problem related to file systems reflects this demand. Additionally, since modern computing is moving towards a more parallel domain, there is a need to understand the bottlenecks of multi-threaded applications~\cite{bandwidthMultithreading} and optimize accordingly. To reflect this, we chose a problem related to false sharing, a typical multi-threading optimization problem~\cite{popularSystemsTricks}. 

We chose problems that fit the following criteria: (1) the problem must have more than one solution where each solution differs not in correctness but runtime performance, (2) The problem should be solvable with or without Copilot assistance in 30 minutes.

\subsection{Problem A: File System Operations}
For Problem A, participants were asked to read records from three large text files. A record is a sequence of 5000 bytes; each file was 1GB. The read operation is specified by \lstinline{FileCombo}, a struct that specifies which file to read from and at what offset. The \lstinline{FileCombo} struct also has a buffer to hold the record read from a file. 

For this problem, participants received the CPP file and the three text files. The full function signatures, the CPP file, and the accompanying documentation given to participants for Problem A are shown in Appendix A.1~\cite{supplement}.

\subsection{Problem B: Multi-threaded Optimization}
For Problem B, participants were asked to use a certain number of threads to set all the values in a source array buffer to zero while setting all the values in a destination array buffer to a particular value. However, they were not allowed to use assignment operations, i.e., move and copy semantics were not allowed on either the source array buffer or the destination array buffer. Participants were only allowed to increment or decrement the values in the respective array buffers. This restriction was in place because we wanted threads to access and modify array items repeatedly, potentially experiencing false sharing. 

The full function signatures, the CPP file, and the accompanying documentation given to participants for Problem B are shown in Appendix A.2~\cite{supplement}.


\section{Model Solutions to the Problems}
\label{section:solutions}
We created \textit{model solutions} to each problem. Because there was more than one solution to each problem, each solution we derived differed only in performance and not correctness. 

We itemize our solutions here and categorize them into Levels 0 -- 3 (L0 -- L3) for Problem A and Levels 0 -- 1 (L0 -- L1) for Problem B. Higher levels correspond to faster runtime performance i.e. L3 has a faster runtime than L0. Details about each of these implementations for both Problem A and Problem B can be found in Appendix B~\cite{supplement}.

\subsection{Problem A Solutions}
\label{subsection:solutions:A}

\subsubsection{Level 0}
\label{subsubsection:solutions:A:L0}
A naive solution to Problem A where calls to \lstinline{open}, \lstinline|seek|, \lstinline|read|, and \lstinline|close| are made for each \lstinline|FileCombo|. 

\subsubsection{Level 1}
\label{subsubsection:solutions:A:L1}
 Using the knowledge that only three files are being interacted with, we do not need to \lstinline{open} and \lstinline{close} a file for each \lstinline|FileCombo|. 
 This optimization involves first opening all the files in \lstinline{FILE_NAMES} and closing them only after all \lstinline|FileCombos| have been processed. This avoids the repeated opening and closing of file descriptors, which is detrimental to runtime performance.

\subsubsection{Level 2}
\label{subsubsection:solutions:A:L2}
Within this optimization, we sort the \lstinline|FileCombos| by \lstinline|fileId| and break ties by \lstinline|offset| before reading the files from storage. As a result, reading records within each specific file will be sequential and not random. Such sequential accesses reduce disk response times, thereby improving program runtime performance.

\subsubsection{Level 3}
\label{subsubsection:solutions:A:L3}
The combination of the L1 and L2 optimizations we outlined above gives us the L3 optimization level, representing the best model solution to Problem A.

\subsection{Problem B Solutions}
\label{subsection:solutions:B}
\subsubsection{Level 0}
\label{subsubsection:solutions:B:L0} Consider a solution to Problem B using \lstinline{THREAD_COUNT} concurrent threads.
A naive solution to this problem is one where all threads start at indices between $0$ and \lstinline{THREAD_COUNT}$-1$ in the \lstinline{src} and \lstinline{dst} arrays. Each thread then decrements and increments one \lstinline{Item} in \lstinline{src} and \lstinline{dst}, respectively. After processing their respective \lstinline{Items}, each thread moves \lstinline{THREAD_COUNT} steps until the next index and processes the \lstinline{Item} therein. For instance, with a \lstinline{THREAD_COUNT} of 4, threads would start at indices $0$--$3$, increment and decrement their respective \lstinline{Items}, before moving 4 steps ahead to their next index.

This is a naive solution because it promotes false sharing. Due to the contiguous nature of the \lstinline{src} and \lstinline{dst} arrays, threads working on \lstinline{Items} with neighboring indices would be operating on the same 64-byte cache lines. As a result, these threads would clash by invalidating each other's cache lines when modifying the \lstinline{Item} within the \lstinline{src} and \lstinline{dst} arrays, leading to \textit{cache thrashing}.

\subsubsection{Level 1}
\label{subsubsection:solutions:B:L1}
False sharing can be avoided by dividing each array (\lstinline{src} and \lstinline{dst}) into \lstinline{THREAD_COUNT} slices and assigning a single thread to process each \lstinline{Item} within a slice. This reduces the probability of mutual cache line invalidation greatly, reducing \textit{cache thrashing}.

Another solution to false sharing would be to add \textit{padding} within the \lstinline{Item} struct definition (See Appendix A.2~\cite{supplement}), bringing its size up to 64 bytes (the cache line size). This would place consecutive \lstinline{Items} within different cache lines, reducing \textit{cache thrashing}. However, we chose not to allow participants to modify the struct definition as this could lead to longer debugging times, potentially violating the time limit constraint for the problem.


\section{Participants}
\label{section:participants}

\begin{figure}
    \centering
    \includegraphics[scale=0.1]{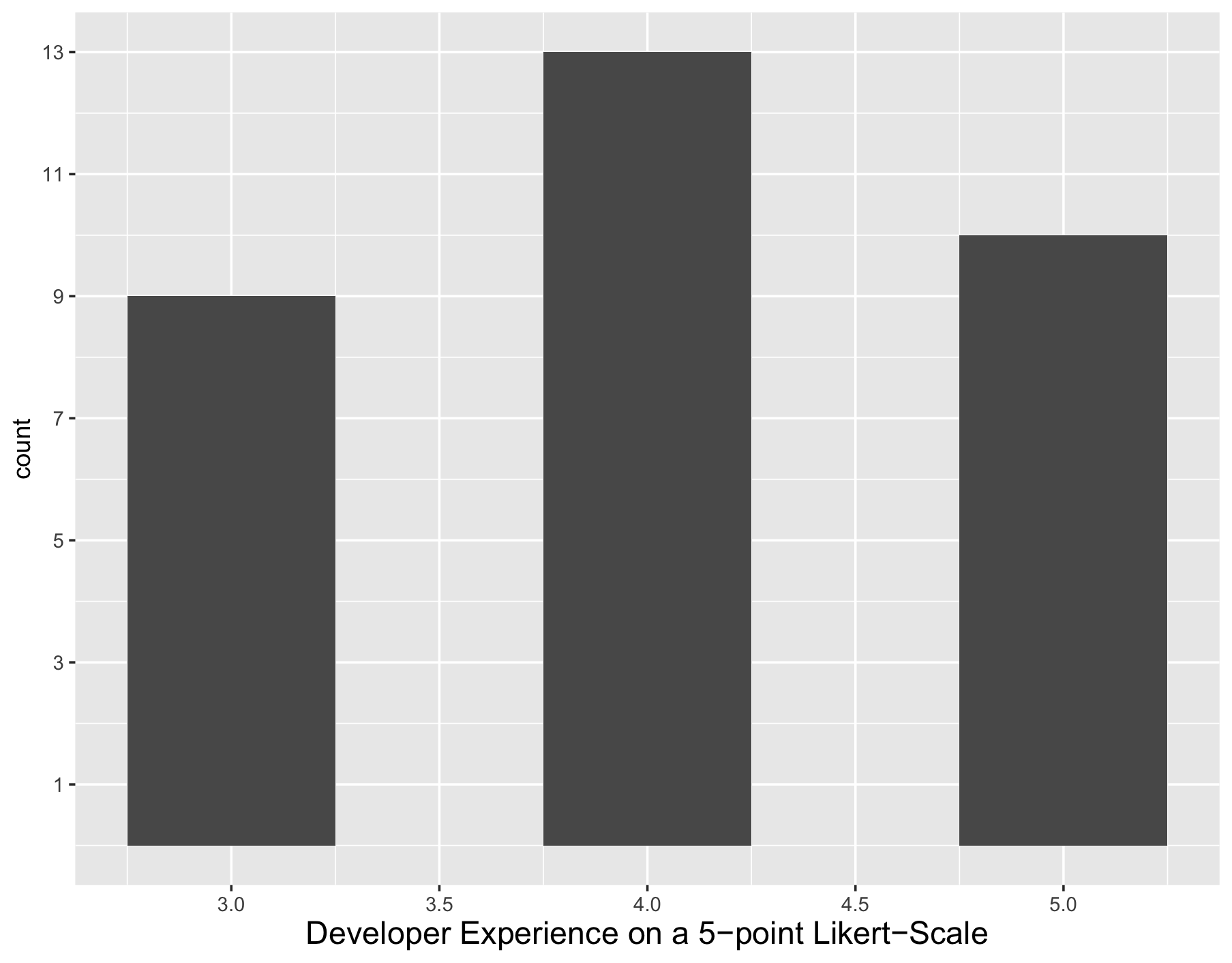}
    \caption{Distribution of Participants' Developer Experience from Screening Survey on a 5-point Likert-Scale from 1 (No experience) to 5 (10 years or more).}
    \label{fig:devExp}
\end{figure}

\subsection{Participant Recruitment}
Participants were recruited mainly via the mailing list for computer science graduate students and snowballed to other interested participants. We primarily targeted participants with experience in systems programming. We considered participants who met one or more of the following conditions to have satisfied this requirement:
\begin{itemize}
    \item The participant has been involved professionally in the Systems / Networking domain, either via industry experience or open-source contributions to systems projects.
    \item The participant has been actively involved in a research project within the Systems / Networking areas.
    \item The participant has taken one or more university courses within the Systems domain including but not limited to Operating Systems, Distributed Systems, or Computer Networking.
\end{itemize}

Additionally, potential participants needed to be familiar with C++, and have access to a web browser as well as GitHub Copilot on Visual Studio Code at the time. Finally, participants could not be employed by OpenAI / GitHub or involved with the development of GitHub Copilot at the time. 

To check if potential participants were eligible to participate, they were sent a Qualtrics screening survey after they signed the consent form declaring their intent to participate. The screening survey can be found in 
Appendix C~\cite{supplement}.

\begin{figure}
    \centering
    \includegraphics[scale=0.1]{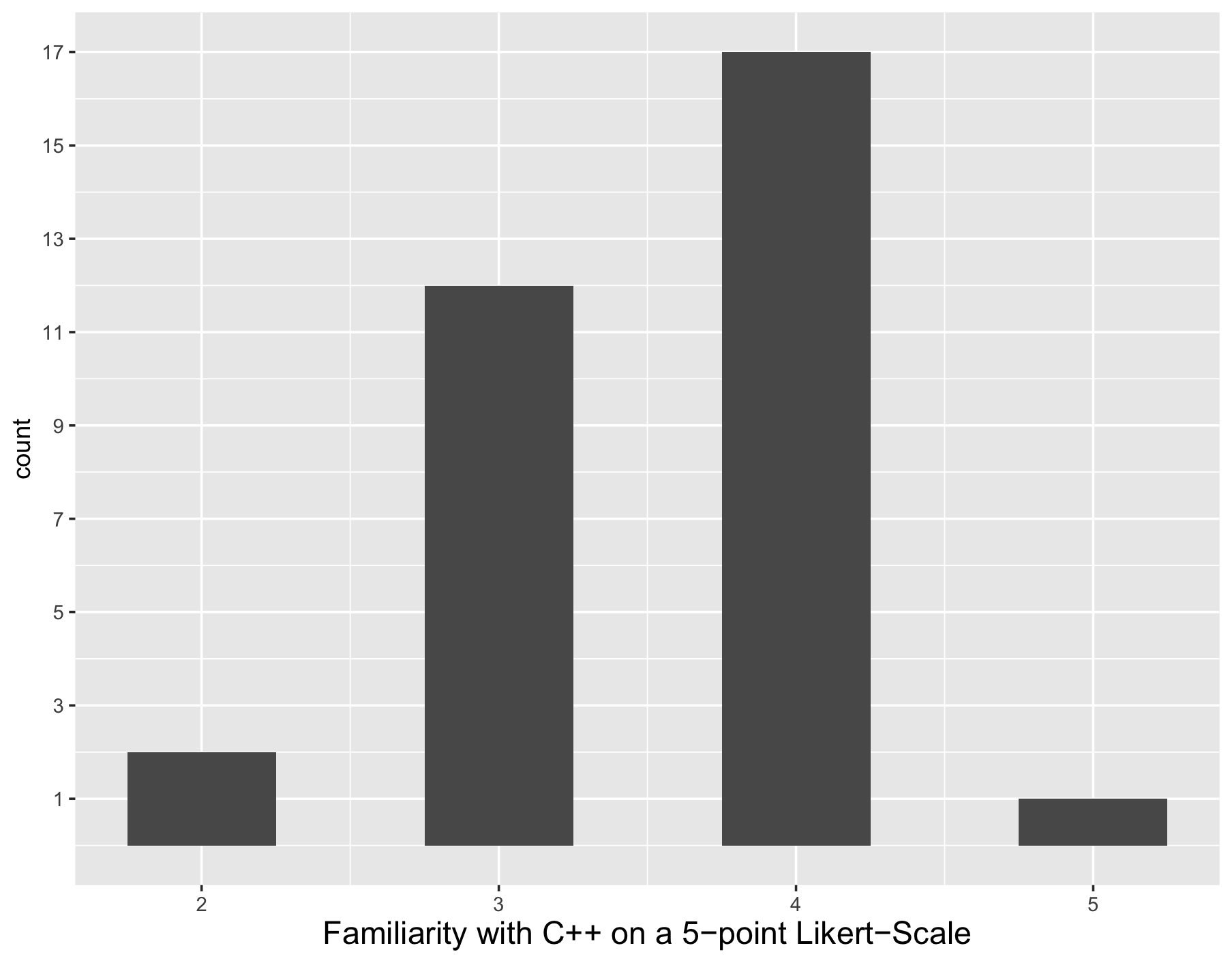}
    \caption{Distribution of Participants' Familiarity with C++ from Screening Survey on a 5-point Likert-Scale from 1 (Not familiar at all) to 5 (Extremely familiar).} 
    \label{fig:familiarCPP}
\end{figure}
\label{difficulty}\subsection{Difficulties Recruiting Professionals}
At the halfway point of our desired participant goal, we paused participant recruitment to analyze the preliminary data collected. 
A majority of the preliminary participants thus far had been graduate students with systems experience, i.e., they were part of systems-focused research groups. We decided to diversify our participant pool by including professional systems developers. 

The initial recruitment process for professional systems developers started with contacting University of Waterloo alumni working within systems-related roles. Additionally, we looked for contributors to open-source systems projects on GitHub which were primarily implemented in C++. The advanced search feature was used to find projects that contained the keywords \texttt{systems}, \texttt{operating systems}, or \texttt{databases}. We also narrowed our search to include only projects with a dedicated social platform where interested parties connect, such as Discord~\cite{discord} and Internet Relay Chat (IRC)~\cite{irc}. 

While projects such as \textit{SerenityOS}~\cite{serenityOS} and \textit{SkiftOS}~\cite{SkiftOS} had active Discord communities, their members were disinterested in the study. Attempts to garner interest within these communities were met with suggestions to reach out to other Discord communities such as the \textit{osdev} (Operating Systems Development)~\cite{osdev} discord channel and the associated IRC. Within the \textit{osdev} communities on Discord and IRC, there was a general unwillingness to participate in the study. Community members cited potential copyright issues with Copilot and other negative perceptions of GitHub Copilot, GitHub, and Microsoft as the primary reasons for their unwillingness to participate in the study. 

However, our persistent recruitment efforts eventually paid off, as we located willing professional participants, enabling us to meet our desired goal.

\subsection{Participant Summary}
We recruited a total of 32 participants for this study, of which 25\% were systems programming professionals or contributors to open-source systems projects. Of the remaining participants, one was a sessional lecturer with systems experience at the University of Waterloo, while the rest were graduate students with a systems research focus. 

Figures ~\ref{fig:devExp} and ~\ref{fig:familiarCPP} show the distribution of participants' experience and their familiarity with C++. Further details about the figures can be found in Appendix C~\cite{supplement}. Participants were compensated \$50 for their time and the study was approved by the Research Ethics Board (\textbf{REB \#44162})
at the affiliated university.


\section{Experiment Design}
\label{section:experimentDesign}

\subsection{Order of Solving the Problems}
Given our within-subjects experimental design where one participant solves one problem with Copilot and then the other problem without it, we needed to ensure that any order effects are counterbalanced across all 32 participants. To this end, we present all the possible orders of the \textit{Problems} (\textbf{A} and \textbf{B}) with the \textit{Modes} (\textbf{C} and \textbf{NC}) which indicate using Copilot and not using Copilot respectively.  The four possible orders of \textit{Mode} $\times$ \textit{Problem} are shown in Table~\ref{table:factorialMatrixOrdered}.

The orders in Table~\ref{table:factorialMatrixOrdered} enforced a requirement that our participant pool be a multiple of four. Hence, we recruited a total of 32 participants for the study.

\begin{table}[h!]
\centering
\begin{tabular}{|l | l | l | l|}
\hline
\textit{\#}       & \textit{First}                & \textit{Second}               & \textit{Participant ID} \\
\hline
1      & C x A    & NC x B    &   P1          \\
2      & C x B    & NC   x A     &   P2          \\
3      & NC x B & C   x A     &   P3          \\
4      & NC x A & C   x B     &   P4          \\

\hline
\end{tabular}
\caption{Possible Orders of \textit{Mode} x \textit{Problem}}
\label{table:factorialMatrixOrdered}
\end{table}

\subsection{Session Overview}

Within this section, we outline the steps carried out in each session. Further details about the tutorial process can be found in Appendix D~\cite{supplement}.

\subsubsection{Pre-session orientation.}~The session was conducted remotely via an online conferencing platform. Each session began with the facilitator introducing the study and confirming the participant's consent to be a part of it. 
After this, screen and audio recording consent for the session was obtained as well. Finally, the facilitator gave the participant a few basic tips for using Copilot such as accepting and rejecting suggestions.

\subsubsection{Session Goals.} Participants were given two C++ programming problems to solve during the session. Each prompt was self contained within a C++ file 
and participants were given a compressed archive containing this file. This compressed archive was sent to the participant via the conferencing platform's chat feature (or Google Drive if technical issues occurred).

The participant was asked to extract the contents of the archive but not open them until the facilitator gave them the signal. After verbally confirming that the participant was ready for the screen capture process to begin, they were asked to share their screen and view the C++ file.

The facilitator then confirmed that (1) all extensions except for the Copilot extension were disabled.\footnote {keybinding related extensions like VSCode Vim~\cite{vscodevim.vim} 
and SSH-related extensions like Remote - SSH~\cite{ms-vscode-remote.remote-ssh} were the only exceptions allowed}(2) the participant could easily switch between their browser and VSCode. The participants were also reminded that the browser and other online resources could be used in addition to GitHub Copilot. 

\label{subsection:experiment-design:timing}
\subsubsection{Timing Constraints.} Before commencement, the participants were notified that they had 30 minutes to tackle each problem. Participants were also alerted at regular intervals such as when 20, 10 and 5 minutes were remaining for each problem.

\label{subsection:experiment-design:after-problem}
\subsubsection{After each problem.} Once the participant declared that they were done with a problem (or the timer ran out), the facilitator stopped the timer and notified the participant. They were then instructed to compress their solution and send it back to the facilitator via the conferencing platform, Google Drive or email.

Once this step was completed, participants were asked to deactivate Copilot (if activated) as well as to close their VSCode window, browser window, and any other references they had opened. This was done to prevent any learning effects that could come from Copilot or the participants (e.g., their browser tabs could contain previous search results or references) from carrying over to the second problem. The participants were sent a link to a survey to complete after which they were allowed a break before tackling the second problem.

The instructions and procedure for the second problem were the same as the first, differing only in the survey at the end. The second survey contained demographic questions in addition to the first survey's questions. Details of the first and second surveys are outlined in Appendix E~\cite{supplement}.

\subsubsection{Post session interview.} At the end of the session, participants were asked for their feedback about the study, GitHub Copilot or anything else they wanted to share.

\section{Evaluation}
\label{section:evaluation}
\textbf{Testbed.} Each participant's code was run on a Linux machine with eight-core Intel Xeon D-1548 at 2.0 GHz, 64GB ECC Memory (4 x 16 GB DDR4-2133), and 256 GB NVMe flash storage. The machine was running Ubuntu 20.04 and the code was compiled with gcc version 9.3.0 \cite{gcc-9}. In order to minimize the effect of small runtime performance variations, we ran each participant's code 32 times with the filesystem cache cleared between each run. 

\textbf{Errors.} If the participant's code did not compile / compiled but encountered runtime errors, it was not analyzed. For instance, one participant's code produced a segmentation fault error even though it compiled successfully. However, if the participants' code compiled, ran without errors but failed the sanity checks, the runtime was recorded but not used in the analysis. As a result, we have only considered correctly implemented solutions when examining the runtime performance.  

\subsection{RQ0 - Does using Copilot influence program correctness?}
Out of our pool of 32 participants, 16 have attempted to solve Problem A with Copilot while the other 16 tackled the problem without its aid. Among the participants who used Copilot for Problem A, every solution passed the sanity checks. On the other hand, among the participants who tackled Problem A without Copilot, 4 out of 16 code snippets either did not compile (P15 and P7), ran and failed the sanity checks (P3), or ran with errors (P23).

Similarly, 16 participants have attempted to solve problem B with Copilot and 16 without its aid. However, in this case, we observed that only 14 code snippets passed the sanity checks both when Copilot was used and when it was not. The 2 ``invalid'' solutions where Copilot was used either did not compile (P15) or ran and failed the sanity checks (P32). On the other hand, the 2 ``invalid'' solutions where Copilot was not used compiled but failed the sanity checks (P30 and P6).

Table~\ref{table:invalid-runs} summarizes these invalid solutions. The fields within the table are described below:

\begin{itemize}
    \item \textit{PartID} - The anonymized ID of the participant
    \item \textit{Problem} - The problem type (\lstinline{A} or \lstinline{B})
    \item \textit{Mode} - Whether Copilot was used (\lstinline{C}) or was not used (\lstinline{NC}) when tackling the problem
    \item \textit{Compiled} - Whether the solution was compiled (\lstinline{TRUE}) or ran into compilation errors (\lstinline{FALSE})
    \item \textit{Passed} - Whether the solution passed sanity checks (\lstinline{TRUE}) or failed them (\lstinline{FALSE}). This field has a value of \lstinline{NULL} if the solution did not compile or ran into runtime errors.
\end{itemize}

\begin{table}[h!]
\centering
\begin{tabular}{|l|l|l|l|l|l|}
\hline
\# & \textit{PartID} & \textit{Problem} &  \textit{Mode} & \textit{Compiled}         &  \textit{Passed}   \\
\hline
1 & P3 &        A &      NC &      TRUE        &  FALSE \\
2 & P7 &        A &     NC &      FALSE         &  NULL    \\
3 & P15 &        A &     NC &      FALSE        &  NULL    \\
4 & P23 &        A &      NC &      TRUE        &  NULL    \\
5 & P15 &        B &     C &      FALSE         &  NULL    \\
6 & P32 &        B &      C &     TRUE          &  FALSE    \\
7 & P6 &        B &      NC &      TRUE         &  FALSE \\
8 & P30 &        B &      NC &      TRUE        &  FALSE  \\

\hline
\end{tabular}
\caption{List of Invalid Runs}
\label{table:invalid-runs}
\end{table}

Our results suggest that \textbf{using Copilot leads developers to produce correct code in most cases}. 

\subsection{RQ1 - Is there a runtime performance difference in C++ code when using GitHub Copilot?}
\label{subsection:evaluation:rq1}
\subsubsection{Approach}
To answer this question, we compare the runtime performance of all 32 runs of the participants' source files for Problems A and B. We use the non-parametric Wilcoxon rank sum test in R~\cite{wilcoxR} \lstinline{wilcox_test()} to compare the runtime performance.

\subsubsection{Results}
\begin{table}[h!]
\centering
\begin{adjustbox}{width=\columnwidth}
\begin{tabular}{|l|l|l|l|l|l|l|}
\hline
\textit{Problem} & \textit{Mode} &  \textit{Valid Runs} & \textit{Mean} & \textit{Median}  &  \textit{Min}  &  \textit{Max} \\
\hline
A &  C &     16 x 32 & 34.86 s &      34.85 s       &  33.82 s    &  36.02 s  \\
A &  NC &    12 x 32 & 26.02 s &      34.47 s       &  4.045 s    &  35.84 s  \\
B &  C &     14 x 32 & 1898 ms &      945.4 ms      &  612.1 ms   &  7356 ms  \\
B &  NC &    14 x 32 & 1628 ms &      943.9 ms      &  494.9 ms   &  6761 ms  \\
\hline
\end{tabular}
\end{adjustbox}
\caption{Summary Statistics of Runtime Performance}
\label{table:summary-statistics}
\end{table}
 
On comparing the runtime performance of valid solutions to Problem A with and without Copilot (\textit{p} = \textbf{3.4e-34}), we find the results to be statistically significant. We observe that solutions without using Copilot were about \textbf{29\%} faster than the ones using Copilot when comparing the mean runtime performance.

Similarly, comparing the runtime performance of the valid solutions to Problem B with and without Copilot (\textit{p} = \textbf{0.000058}), we also find the results to be statistically significant. Again, we observe that solutions without using Copilot were about \textbf{15\%} faster than the ones using Copilot when comparing the mean runtime performance.

\mei{Table~\ref{table:summary-statistics} highlights the summary statistics of the runtime performance for participants' valid solutions to the problems. From this we can see that while the mean runtime performance is quite different, the median runtime performance in both problems are closer when comparing solutions created with and without Copilot. Even though the values are closer, not using Copilot still has a marginally faster runtime than using Copilot. This observation from medians along with the min and max values tell us that there are outliers in the data. These outliers matter too. 

We notice that the fastest solution to Problem A is when not using Copilot and is 8 times faster than the median solution to the same problem with or without Copilot. However the same participant who wrote the fastest code for Problem A without Copilot had an average runtime performance of approximately 905 ms for Problem B when using Copilot. This value is much closer to the median as we can see from Table~\ref{table:summary-statistics}. Thus, we can see that the same participant when using Copilot did not write the same high performance code. Note also that the max times are always faster when not using Copilot than when using Copilot. From all these comparisons and the statistical testing, we can see a picture emerging where participants who used Copilot always wrote code that has slower runtime performance than than those who did not.}

For further context into the runtime performance, we also ran our L1, L2, and L3 solutions to Problem A and our L1 solution to Problem B for 32 runs alongside the participants' solutions. In Table~\ref{table:model-solutions-runtime} we see that our L1 solution to Problem A was \textbf{13\%} faster and \textbf{16\%} slower than participants' Copilot-aided and Copilot-unaided solutions respectively. Our L2 solution to Problem A was \textbf{129\%} and \textbf{110\%} faster than participants' Copilot-aided and Copilot-unaided solutions, respectively. Our L3 solution to Problem A was \textbf{147\%} and \textbf{132\%} faster than participants' Copilot-aided and Copilot-unaided solutions, respectively.

Similarly, our L1 solution to Problem B was \textbf{106\%} and \textbf{95\%} faster than participants' Copilot-aided and Copilot-unaided solutions. We did not run our L0 solutions because the participants already implement L0 solutions for both problems

\begin{table}[h!]
\centering
\begin{tabular}{|l|l|l|}
\hline
\textit{Problem} & \textit{Level} &  \textit{Mean} \\
\hline
A &  L1 &     30.59 s   \\
A &  L2 &    7.565 s    \\
A &  L3 &     5.228 s   \\
B &  L1 &    581.4 ms    \\
\hline
\end{tabular}
\caption{Model Solutions Runtime Performance}
\label{table:model-solutions-runtime}
\end{table}

\subsubsection{Discussion}
Our results suggest that \textbf{developers may benefit from Copilot-unaided C++ code in terms of runtime performance}. We give further context to these results by highlighting some participants' Copilot-unaided solutions whose mean runtime performance was close to or better than the model solutions highlighted in Section~\ref{subsection:solutions:A} and Section~\ref{subsection:solutions:B}.

\textbf{Problem A.} While our model L3 solution had a mean runtime of 5.288 s, P31's noteworthy Copilot-unaided solution had a mean runtime of 4.547 s beating our best model solution by \textbf{15\%}. Their solution is shown in Listing~\ref{listing:P31:L3:A:NC}. 

We note that their solution used the L3 optimization for Problem A discussed in Section~\ref{subsubsection:solutions:A:L3}. Additionally, in lines 4 - 7 a map was used to associate each \lstinline|fileId| with a vector of \lstinline|fileCombos| for the associated file. The pre-processing in this step allowed them to sort each vector of \lstinline|fileCombos| belonging to a file (line 9), open the file once (lines 11 - 12), process all the \lstinline|fileCombos| (lines 13 - 16) and then close the file (line 17). While the fundamental concept of the L3 optimization is still present, some implementation details are slightly different and as such may have contributed to the observed speed-up. 

It is also pertinent to mention that P31 had ideas to add other optimizations that could have potentially reduced the runtime performance of their code even further. However, they did not have sufficient time to do so and debug their solution. They outlined this optimization in code comments which have been removed from the Listing for clarity. The potential improvement involved the usage of \lstinline{memcpy}~\cite{memcpy} ``to avoid overlaps''. 

\begin{listing}[h!]
\cppcode{listings/listing-P31-L3-a-NC.cpp}
\caption{P31’s L3 Solution to Problem A without Copilot}
\label{listing:P31:L3:A:NC}
\end{listing}

\textbf{Problem B.} A noteworthy solution to Problem B was P17's Copilot-unaided solution (in Appendix F~\cite{supplement}). This resembled the model L1 solution with some statement-level optimizations explained in Section~\ref{subsection:evaluation:rq2} and was one of the closest-performing solutions to our L1. Their solution had a mean runtime performance of 636.4 ms which was only 9\% slower compared to our model L1 solution, which had a mean runtime performance of 581.4 ms.

\subsection{RQ2 - Do Copilot’s suggestions sway developers towards or away from C++ code with faster runtime performance?}
\label{subsection:evaluation:rq2}

\subsubsection{Approach}
We wanted to understand how suggestions from Copilot swayed participants to produce code with slower or faster runtime performance. To this end, we took the last snapshot of the participants' submitted code and categorized each participant's code for problems A and B. We labelled participants' code according to the optimizations discussed in Section~\ref{section:solutions}. 

An author of this work and a collaborator separately looked through the source code for all participants and labelled each solution for Problem A as either L0, L1, L2, or L3 to indicate the levels of optimizations that participants used. Similarly, for Problem B, they were labelled as L0 or L1. Additionally, they also noted programming constructs that participants used that could potentially increase or decrease the runtime performance and tried to group similar constructs.

We term these ``programming constructs'' as statement-level optimizations and refer to the optimizations within Section~\ref{section:solutions} as concept-level optimizations from this point.

\subsubsection{Statement Level Optimizations \& Open-coding}
After the author and the collaborator finished labelling participants' source files with concept-level and statement-level optimizations, they came together to resolve disagreements and discuss emerging patterns in the statement-level optimizations and remarks. Upon resolving the disagreements, they came up with a set of themes to encompass the statement-level optimizations. A summary of these themes/categories of statement-level optimizations for Problem A and Problem B are in Table~\ref{table:optimizationsCodingA} and Table~\ref{table:optimizationsCodingB} respectively.

\subsubsection{Video Analysis}
\label{subsubsection:evaluation:rq2:videoAnalsis}
Using the themes generated in Table~\ref{table:optimizationsCodingA} and Table~\ref{table:optimizationsCodingB}, the author went through all 32 screen-shared recordings of participants solving the problem when Copilot was used and tracked the accepted suggestions or series of accepted suggestions that participants accepted that swayed them to the solutions that fit their themes. 

\subsubsection{Results}
For Problem A, where Copilot was used, 15 of the 16 correct solutions used the L0 naive implementation with the \lstinline{<fstream>}~\cite{fstream} family of library functions and thus were categorized as L0F. Additionally, few remarks were made as most solutions only used the naive L0F implementation in \ref{subsubsection:solutions:A:L0}. Some solutions were remarked as NCLOSE because they failed to close the files after reading from them. Some solutions also landed in the BINARY category. From the video analysis, it would seem that Copilot largely gave L0F suggestions, and participants simply accepted them without editing. Participants also only confirmed that the sanity checks passed before declaring they were done with the problem.

In Problem B, we notice a relatively balanced use of concept-level optimizations and varied use of statement-level optimizations and remarks when using Copilot. From 14 (out of 16) source snippets with correct solutions, we note that 9 of those solutions used the L1 concept-level optimizations of avoiding false sharing. Notably, 1 of the 9 (P23) was classified as L1 because it avoided false sharing by using OpenMP to handle the multi-threaded execution. 2 of the 14 solutions were encoded as L0 even though false sharing was absent because they either used a single-threaded approach (P7) or used only one additional thread (P3) for the problem instead of \lstinline{THREAD_COUNT} threads. 3 of the 14 (P4, P11 and P19) solutions were encoded as L0 because false sharing was present in their solutions. Additionally, statement-level remarks such as 2LOOPS or 1LOOP were prevalent in the solutions.

Moreover, ITER\_NAIVE and ITER\_FAST were also common categories that emerged. Rarer categories like OPENMP, ONET and NT also appeared in a few cases. From the video analysis, Copilot initially suggested incomplete snippets leaning toward L0. Participants would accept the snippets and try to get the rest of the solution to work by debugging. In other cases, participants wrote comments about dividing an array into  \lstinline{THREAD_COUNT} chunks, and Copilot would suggest snippets leaning towards L1.

\subsubsection{Discussion}
For Problem A with Copilot, there was an interesting case where P22 was swayed via Copilot's suggestions to use L1U (Level 1 optimization but using the \lstinline{<unistd.h>}~\cite{unistd} and \lstinline{<fcntl.h>}~\cite{fcntl} I/O functions). From the video analysis, we observe that the participant was largely responsible for coming up with concept-level L1 optimization in that they only declared a vector of file descriptors before the suggestions to use L1U with NCLOSE came along, which the participant accepted. However, P22 remarked that they ``had to do more post-hoc checking'' instead of ``figuring out how to solve the problems''; that it was ``a different approach of how they would solve the problem''. We also note that while their solution used the L1 concept-level optimization, the mean runtime for their solution was \textbf{35.48 s} which was \textbf{15\%} slower than our model L1 solution. This difference may be due to differences in the I/O implementation details in the \lstinline{<unistd.h>} and the \lstinline{<fcntl.h>} libraries versus the \lstinline{<fstream>}~\cite{fstream} library. A snippet of P22's Copilot-aided solution to Problem A can be found in Appendix F~\cite{supplement}.

Within Copilot-aided solutions to Problem B, we noticed that the solution with the least mean runtime performance at \textbf{677.8 ms} was from P12, who used the L1 concept-level optimization, and landed in the 1LOOP and ITER\_LESS\_NAIVE themes for the statement-level remarks. From the video analysis, the initial incomplete solutions accepted by the participant were leaning towards 1LOOP, NT and the incorrect solution of MISSING\_LOOP. P12 was primarily responsible for implementing the code in the ITER\_LESS\_NAIVE statement-level remark because they ``didn't think Copilot understood them[me] well when they[I] told it to increment or decrement'' and ``just gave up and wrote it themself[myself]''. However, the L1 suggestion to split the thread into slices was accepted by the participant without much editing. P12 also remarked that ``Copilot was useful'', and they ``usually just google'' what Copilot would have suggested. We also note that their solution was 16\% slower than our model L1 solution which could be because the model L1 solution used ITER\_FAST and 1LOOP statement level optimizations. See a snippet of P12's solutions to Problem B that was done with Copilot in Appendix F~\cite{supplement}.

Some interesting categories for statement level optimizations in Problem B in Table~\ref{table:optimizationsCodingB} are worth taking a closer look at, notably, 2LOOPS, 1LOOP and ITER\_NAIVE and ITER\_FAST. Our model L1 solution uses 1LOOP, ITER\_FAST and also avoids false sharing and averages at a mean of \textbf{581.4 ms}. The closest Copilot-aided solution to the model solution in terms of runtime performance was P12's (Appendix F~\cite{supplement}) with a mean runtime performance of \textbf{677.8 ms}. At a close second was P24 (Appendix F~\cite{supplement}) with a mean runtime performance of \textbf{784.0 ms}, which avoided false sharing and used 1LOOP and ITER\_NAIVE. This difference in runtime performance between the model L1 solution and P24's suggests that using ITER\_FAST is better than using ITER\_NAIVE to update the source and destination buffers when false sharing is avoided. If we also look at P27's Copilot-aided solution to Problem B (See Appendix F~\cite{supplement}), we notice that while it avoids false-sharing, it uses 2LOOPS and ITER\_NAIVE which earns it a mean runtime performance of \textbf{925.1 ms}. Comparing P24's with P27's solution suggests that using 2LOOPS instead of 1LOOP to update the source and destination buffers when false sharing is avoided could result in slower runtime performance. On the other hand, if we look at solutions where false sharing was used, we note that both P11's (See Appendix F~\cite{supplement}) and P19's (See Appendix F~\cite{supplement}) Copilot-aided solutions had false sharing present. However, their solutions used 2LOOPS with ITER\_NAIVE with a mean runtime performance of \textbf{1434 ms} and 1LOOP with ITER\_NAIVE with a mean running time of \textbf{6202 ms}, respectively. This difference in runtime performance may suggest that using 1LOOP instead of 2LOOPS could result in slower runtime performance when false sharing is present, which is different from when false sharing is absent, as with P24's and P27's solutions.

\begin{table*}[h!]
\begin{adjustbox}{width=\textwidth}
\centering
\begin{tabular}{|l|l|}
\hline
Encoding & Summary \\
\hline
L*F        & Used \lstinline|<fstream>| \cite{fstream} library for any of the concept-level optimizations L0, L1, L2, or L3                                                                                  \\
L*C        & Used \lstinline|<cstdio>| \cite{cstdio} library for any of the concept-level optimizations L0, L1, L2, or L3                                                                                   \\
L*U       & Used \lstinline|<unistd.h>| \cite{unistd} and \lstinline|<fcntl.h>| \cite{fcntl} libraries for any of the concept-level optimizations L0, L1, L2, or L3                                                                \\
NCLOSE     & Did not close file                                                                                                                   \\
EXCEPT     & Added \lstinline|file.exceptions(...)| \cite{exceptions} to catch possible exceptions                                                              \\
ASSERT     & Asserted that no error flags were set after file operations using 
\lstinline|good()| \cite{good} method and other assertions to ensure program correctness \\
READ\_COMBO & Helper function for processing a single \lstinline|fileCombo| in \lstinline|fileCombos| and by calling \lstinline|open()|, \lstinline|seek()|, \lstinline|read()|, and \lstinline|close()| in order         \\
BEGIN      & Explicit seek from \lstinline|std::ios_base::beg| \cite{seekdir} in call to \lstinline|seekg()| \cite{seekg}                                                                               \\
OC\_WITHIN  & Opened and closed the files within the same loop as processing each \lstinline|fileCombo| in \lstinline|fileCombos|                                                          \\
BINARY     & Added a "binary" flag to the \lstinline|open| call using \lstinline|std::ios::binary| \cite{openmode} or similar                                                         \\
MAP        & Used a \lstinline|map| \cite{map} to associate a file with all the \lstinline|fileCombos| associated with that file      \\                                         
\hline
\end{tabular}
\end{adjustbox}
\caption{Table of Statement-level Optimizations \& remarks for Problem A}
\label{table:optimizationsCodingA}
\end{table*}
\begin{table*}[h!]
\begin{adjustbox}{width=\textwidth}
\begin{tabular}{|l|l|}
\hline
Encoding & Summary \\
\hline

NT                & No threads used                                                                                                                                                  \\
ONET              & Only one thread was used. Equivalent to not using threads                                                                                         \\
MISSING\_LOOP     & Failed to loop to decrement \lstinline|src[i]| to zero and to increment \lstinline|dst[i]| to \lstinline|INIT_SRC_VAL|. This is an incorrect solution.       \\                      
ITER\_NAIVE       & Made \lstinline|SIZE X INIT_SRC_VAL| repeated calls to \lstinline|dst[i].get()| or \lstinline|src[i].get()| while decrementing \lstinline|src[i]| and incrementing \lstinline|dst[i]|                                       \\
ITER\_LESS\_NAIVE & Made \lstinline|SIZE| repeated calls to \lstinline|src[i].get()| or \lstinline|dst[i].get()| while decrementing \lstinline|src[i]| and incrementing \lstinline|dst[i]|               \\
ITER\_FAST        & No calls to \lstinline|src[i].get()| or \lstinline|dst[i].get()| while decrementing \lstinline|src[i]| and incrementing \lstinline|dst[i]| but iterated up to \lstinline|INIT_SRC_VAL|                                                                                            \\
2LOOPS            & Decremented \lstinline|src[i]| to 0 then incremented \lstinline|dst[i]| to \lstinline|INIT_SRC_VAL| instead of in lockstep                                                                    \\
1LOOP             & Decremented \lstinline|src[i]| and incremented \lstinline|dst[i]| in lockstep                                                                                                        \\
SPLIT             & \lstinline|src[i]| is decremented using a separate thread and \lstinline|dst[i]| is incremented using a separate thread                                                              \\
SPLIT2            & Like SPLIT but \lstinline|src[i]| decremented using 2 threads after being divided into 2 slices and \lstinline|dst[i]| incremented using 2 threads after being divided into 2 slices \\
MANY\_SPLIT       & Spawned \lstinline|SIZE| threads where each thread handled \lstinline|src[i]| and \lstinline|dst[i]|. There could be context switches since not enough threads on machine                      \\
LOCKS             & Used locks.                                                                                                                                                      \\
RACET             & Race conditions in thread spawning without locks leading to incorrect results                                                                                    \\
HARDT             & Hardcoded thread spawning instead of dynamic based on \lstinline|THREAD_COUNT|                                                                                                            \\
PTHREAD           & Used \lstinline|pthread_create| and \lstinline|pthread_join| \cite{pthread} to create and join threads instead of \lstinline|std::thread| methods                              \\
SPAWN\_SEP        & Spawned \lstinline|THREAD_COUNT| threads to process \lstinline|src[i]| then wait to finish then spawn another \lstinline|THREAD_COUNT| threads for \lstinline|dst[i]| then wait to finish              \\
OPENMP            & Used \lstinline|parallel for| in OpenMP.    \\                                                                                 

\hline
\end{tabular}
\end{adjustbox}
\caption{Table of Statement-level Optimizations \& remarks for Problem B}
\label{table:optimizationsCodingB}
\end{table*}



\added{\section[discussion]{Discussion}
\label{section:discussion}

\subsection{Comparison of results between students and professionals}

To see if students and professionals who took part in our study had different outcomes, we split the data we had between students and professionals. Then we compare the runtime performance of the solutions written by students and professionals with and without Copilot for problems A and B. We compare the results using box plots and carry out the non-parametric
Wilcoxon rank sum test for statistical significance. 

\noindent\textbf{Problem A:} From Figure \ref{fig:5} we can see the comparative run time performance of the solutions from students and professionals in Problem A. In both Figures. \ref{fig:5a} and \ref{fig:5b} we can see that the runtime performance of solutions when both students and professionals don't use Copilot is faster than when they use Copilot. The difference in the mean runtimes is approximately  8-9 seconds slower when using Copilot. When we test the data using the non-parametric Wilcoxon rank sum test, we find that the results are statistically significant (p = 4.99e-22 and p = 2.76e-15 for students and professionals respectively). Hence we can conclude that in Problem A, when taken as a whole and separately (as students and professionals), we consistently get the result that C++ code written with Copilot is slower than C++ code written without Copilot. 

\noindent\textbf{Problem B:} In Figure \ref{fig:6}, we can boxplots of the runtimes between students and professionals when they solve problem B with and without Copilot. We can again see that the mean runtimes when using Copilot is slower than when not using Copilot for both students (Figure \ref{fig:6a}) and professionals (Figure \ref{fig:6b}). The difference in means is approximately 0.2-0.5 seconds. When we test the data using the non-parametric
Wilcoxon rank sum test, we find that the results are statistically significant for professionals (p = 2.61e-9) and not for students (p = 0.049). Therefore, while we cannot statistically conclusively say that the runtimes are slower for both students and professionals when considered separately,  we see that the relationship between using and not using Copilot for Problem B is consistent with the data as a whole.  

\begin{figure}
\begin{subfigure}[b]{0.25\textwidth}
   \includegraphics[width=\linewidth]{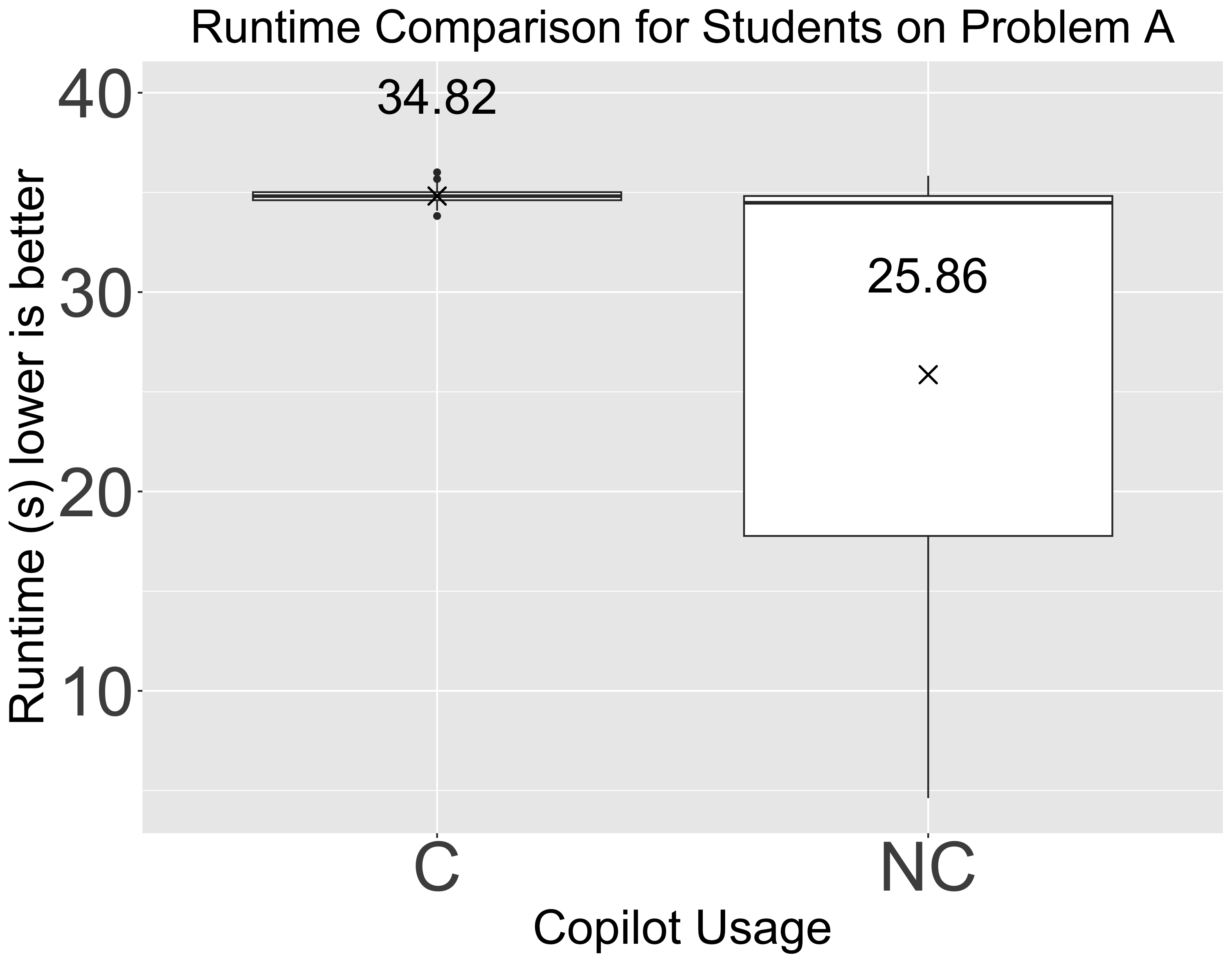}
    \caption{Students} \label{fig:5a}
  \end{subfigure}%
\begin{subfigure}[b]{0.25\textwidth}
   \includegraphics[width=\linewidth]{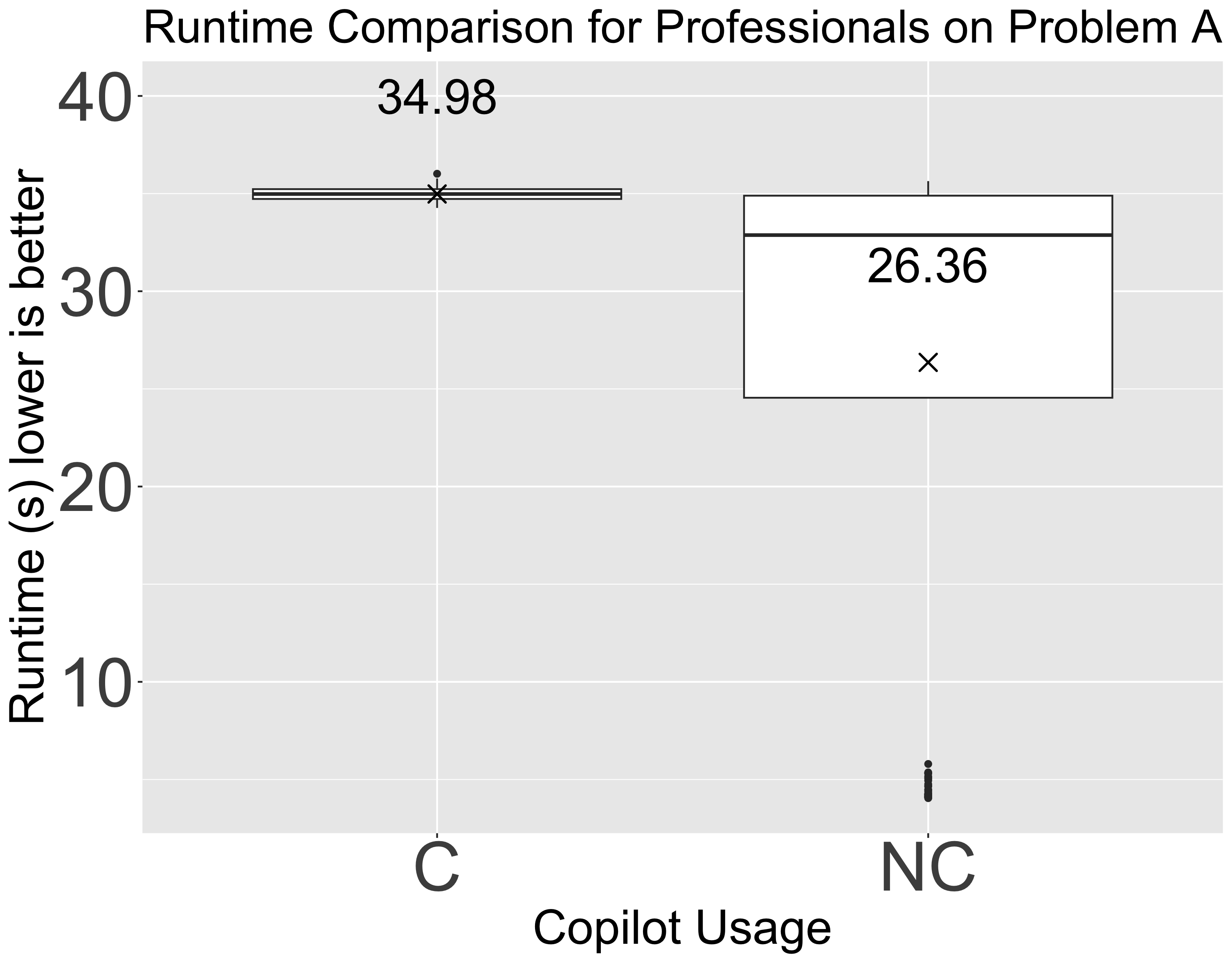}
    \caption{Professionals} \label{fig:5b}
  \end{subfigure}%
\caption{\added{Box Plots of runtimes for solving Problem A with Copilot (C) and Without Copilot (NC)}} \label{fig:5}
\end{figure}

\begin{figure}
\begin{subfigure}[b]{0.25\textwidth}
   \includegraphics[width=\linewidth]{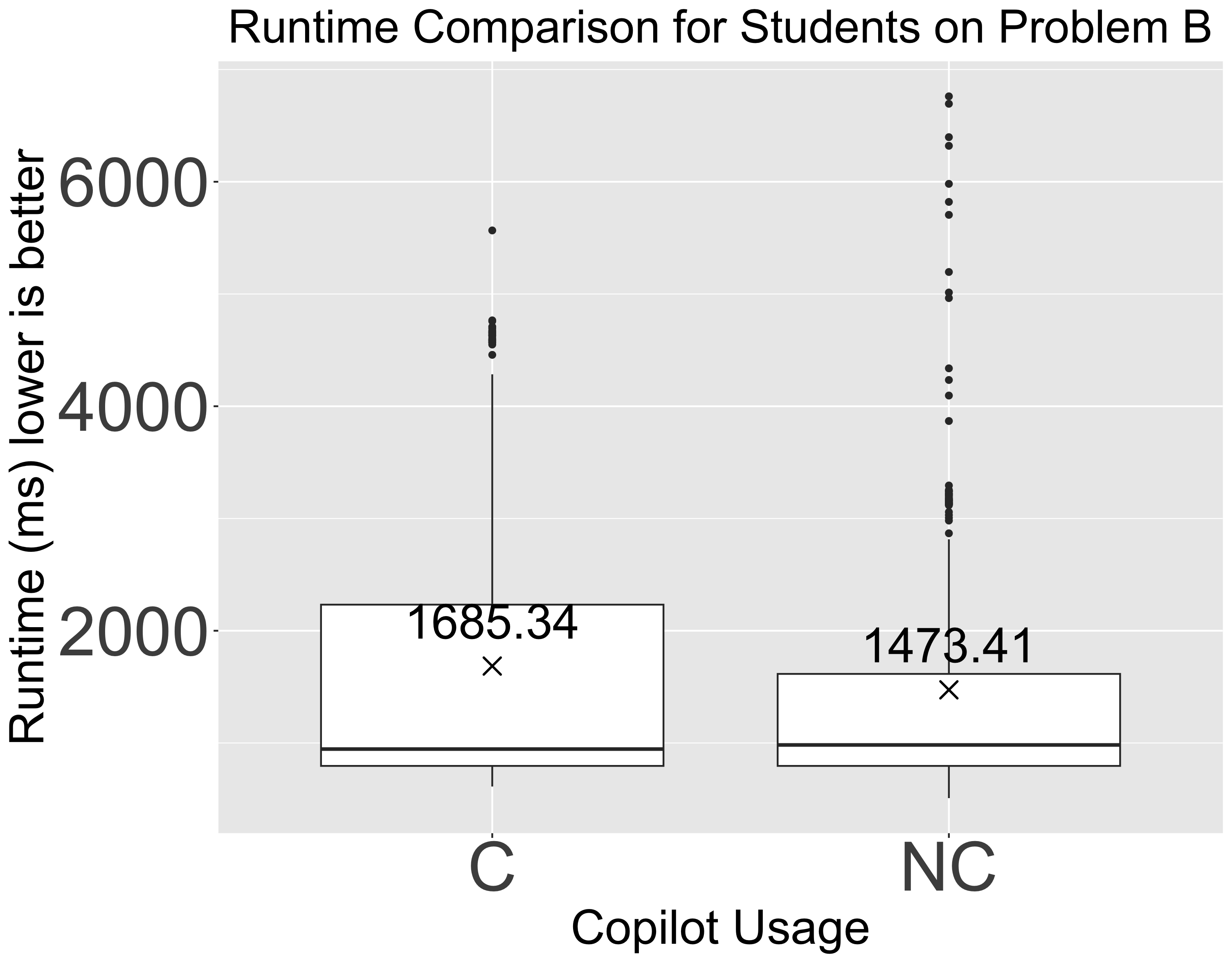}
    \caption{Students} \label{fig:6a}
  \end{subfigure}%
\begin{subfigure}[b]{0.25\textwidth}
   \includegraphics[width=1\linewidth]{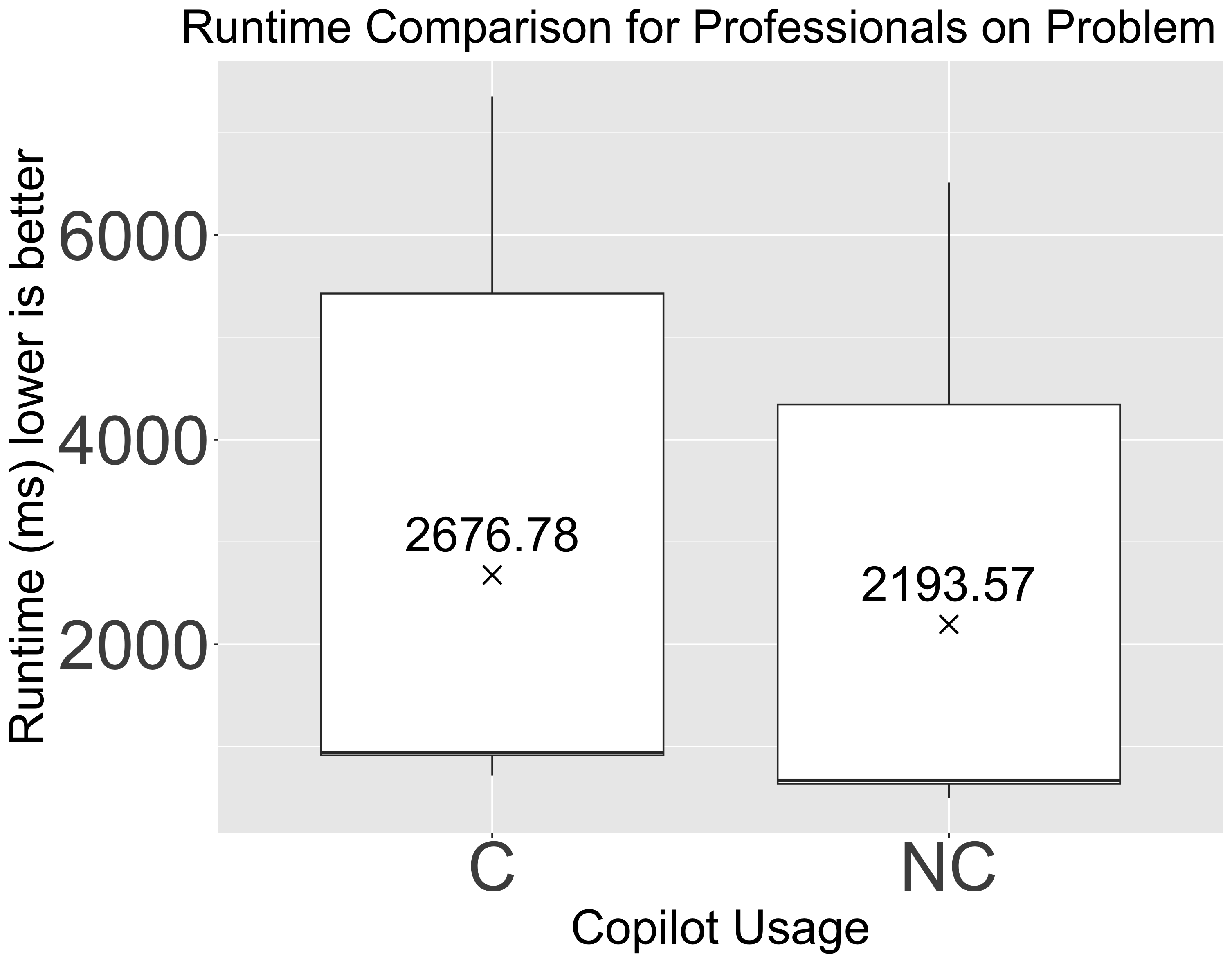}
    \caption{Professionals} \label{fig:6b}
  \end{subfigure}%
\caption{\added{Box Plots of runtimes for solving Problem B with Copilot (C) and Without Copilot (NC)}} \label{fig:6}
\end{figure}

From Figures \ref{fig:5} and \ref{fig:6} we can conclude that our results from Section~\ref{subsection:evaluation:rq1} still hold when we consider students and professionals separately for both problems A and B. 
\subsection{Familiarity with the problem}

Two of the questions we asked participants in the post-survey were whether they had previously seen the problem that they solved, and if they had solved it when previously seen. Out of the 64 combinations (32 participants each solving 2 problems, one with Copilot and one without = 32x2), we found that only in 8 cases had participants either seen or solved the problem before~(see Table ~\ref{table:familiarity}. Therefore an overwhelming majority of them had neither seen the problem nor solved it before. 

\begin{table}[h!]
\centering
\begin{adjustbox}{width=\columnwidth}
\begin{tabular}{|l|l|p{1cm}|p{1cm}|p{1cm}|l|p{1cm}|}
\hline
\textit{Problem} & \textit{Mode} &  \textit{Have Seen Problem} & \textit{Have Solved Problem} & \textit{Is a Professional}  &  \textit{Runtime}  &  \textit{Difference With Mean of the Setting (D)} \\
\hline
A	&	C	&	Yes	&	Yes	&	TRUE	&	35.01	&	0.03	\\
A	&	NC	&	Yes	&	Maybe	&	FALSE	&	21.53	&	-4.33	\\
A	&	NC	&	Yes	&	No	&	FALSE	&	34.95	&	9.09	\\
B	&	C	&	Yes	&	Yes	&	FALSE	&	773.99	&	-911.35	\\
B	&	C	&	Yes	&	No	&	FALSE	&	1433.83	&	-251.51	\\
B	&	C	&	Yes	&	No	&	FALSE	&	3561.19	&	1875.85	\\
B	&	C	&	No	&	Yes	&	FALSE	&	2245.17	&	559.83	\\
B	&	NC	&	Yes	&	No	&	FALSE	&	1404.06	&	-69.35	\\
\hline
\end{tabular}
\end{adjustbox}
\caption{\added{Participant familiarity with the problem and the solution. Column D at the end notes the difference between the runtime of the solution from a participant with the mean of the runtimes from all participants. A negative value indicates that the participant had a faster solution than the mean.}}
\label{table:familiarity}
\end{table}

In addition from Table~\ref{table:familiarity} we see that in 4 cases participants had a faster solution and in 4 cases they had slower solutions compared to the mean runtime for that setting (Problem x Mode). Interestingly, in problem A we have a professional who has both seen and solved the problem before. They used Copilot to solve Problem A in our experiments and produced a solution with a slightly slower runtime than the mean runtimes when using Copilot to solve Problem A. This indicates that even when people have solved the problems, Copilot may lead them to a slower-than-average solution. On the other hand, students using Copilot to solve Problem B were evenly split. Two of them had a faster solution and two had a slower solution compared to the mean.  And when the student in the last row on Table~\ref{table:familiarity} solved Problem B without Copilot, they had a faster runtime than the mean. 

From the results in this analysis, we can see that (a) our experiments were not biased much with results from people who had seen or solved the problem before, and (b) even when they have, the results indicate that using Copilot may result in solutions with slower performance than when not. 
}

\added{
\section[design]{Threats to Validity}
\label{section:rational}
\subsection{External Validity}
\textbf{Programming Language and Code Generation Tool.} This study is explicitly about the runtime performance of C++ code written with the help of Copilot. We specifically chose C++ as a programming language for the experiment as C++ applications are typically performance-critical. We also specifically chose Copilot as it is used by more than a million developers~\cite{copilot_marketshare}. We need more studies to examine runtime performance in other settings - different programming languages and different LLM-based code generation tools. 

\mei{\textbf{Number of Problems:} We restrict our study to two problems as increasing the number of problems increases the number of participants required exponentially in our controlled experiment. To maintain the same experimental design we would need 384 participants for 4 problems and 11520 participants for 6 problems. The effort to run the experiment with more than 12-360 times the current number of participants is exponential in many ways: time to find participants, run the experiment, pay the participants, and analyze the data from the experiment. We do not know of any software engineering research study with a controlled experiment where there were more than 300 participants. There have been survey-based studies with more than 300 participants, but surveys are not high in effort like a controlled experiment. A recent paper using LLMs for code understanding also has 32 participants and 2 tasks~\cite{nam2024using}.}

\mei{\textbf{Choice of Problems:} We acknowledge the limitations of the representativeness of the programming problems used for this study. While file system operations and multi-threading programming concepts are critical, there could be other important domains that are not represented in our study that developers could have been more aware of. However, we argue that our chosen problem domains are typically part of the training of software engineers making it worth examining.}


\subsection{Construct Validity}
\mei{\textbf{Not explicitly asking participants to write low runtime performance code:} Although it might seem that explicitly directing participants to produce performant code with or without Copilot, we argue that this directive would be part of a different study. This future study would answer the question of how well Copilot can generate highly performant code compared to human developers. }
\mei{Typically while run time performance is desired, we have not seen a case where every requirement in a software is explicitly asked to have lower runtime performance. Hence, our study explores whether code written without any explicit requirement for performance is different when using Copilot. Another possible study is when compared to developers with no performance experience, can directed prompts in Copilot produce higher-performance code. The participant pool, experiment designs and analysis of all of these studies are quite different. We leave these alternate studies as future work that needs to be studied too and argue that our study and its results are valuable too. }

\mei{\textbf{Choice of participants:} 
We acknowledge that 75\% of our participants are graduate students. We intended to have more than our current set of professionals~(25\%). However, as we state in Section~\ref{difficulty}, we faced difficulty in recruiting professionals, especially systems programmers. However, all the graduate students who did participate are systems researchers who have extensive C++ and developer experience. Additionally, we split the results for students and professionals, and found that the findings remained the same - students and professionals wrote code that on average had a faster runtime performance when not using Copilot in comparison with using Copilot for both problems A and B. We have included the tables in our replication dataset online~\cite{supplement}.}

\mei{Finally, we acknowledge that there could have been some participant selection bias as we only selected participants that wanted to participate in the study. Our results may have been slightly different if the developers that were unwilling to participate actually took part in the study. A workaround to recruit such developers unwilling to participate due to negative perceptions about GitHub Copilot would be to omit details about using it until the session actually began. However, as there are significant ethical concerns with this type of deception in controlled human studies, this was infeasible.}

\subsection{Internal validity}
As we were looking at the runtime performance of participants' code, another possible limitation could have been that participants did not have enough time to optimize their solution. However, on average all 32 participants spent approximately \textbf{17 minutes} of the 30 minutes allotted on problem A and \textbf{20 minutes} on problem B. Therefore the participants were satisfied with their solution at least 10 minutes before their time was up. 
}

\added{
\section[Conclusion]{Conclusions and Takeaways}
\label{section:conclusion}


This work evaluated the performance of C++ code generated by the self-proclaimed AI programming assistant GitHub Copilot by conducting a user study on systems programmers. We present our main takeaways for different stakeholders.
\begin{itemize}
    \item{\textbf{Developers:}} Developers must be careful about the code they get from Copilot, and not only review Copilot-generated code for functional correctness but also for non-functional aspects like runtime performance.
    \item{\textbf{Maintainers:}} Maintainers of Copilot need to evaluate and improve their tools and models to focus not only on functional correctness but also on other aspects of code that are just as important: performance, security, reliability, etc.
    \item{\textbf{Researchers:}} While benchmark datasets like HumanEval, MBPP, and SWE-Bench are available to evaluate the functional correctness, we need benchmark datasets for non-functional aspects too. With GitHub Copilot gaining ubiquity in modern software development, more research is required to scope its strengths and limitations.
\end{itemize}

}

\section{Data Availability}
We provide the problems and prompts given to participants, the expert solutions we generated, the test scripts we used for evaluation, and the data set of the runtime performance of participants' solutions as well as the runtime performance of the model solutions~\cite{supplement}. \mei{However, to respect participant privacy and anonymity as well as to abide by the rules set by the ethics review board, we are unable to share participant responses to the screening survey, their video data, their responses to the programming surveys, and their entire unedited source code solutions. We share everything needed for anyone to be able to replicate this study with their own set of participants by ACM standards~\cite{acmreplication}. }

\ifCLASSOPTIONcaptionsoff
  \newpage
\fi



\bibliographystyle{IEEEtran}
\bibliography{references}
%

\end{document}